\documentclass{WileyChemistry-template}

\usepackage{amsmath,amssymb}%
\usepackage{dsfont}%
\usepackage{amsfonts}%
\usepackage{amsthm}%

\makeatletter
\newcommand{\@defs@vec}[1]{\boldsymbol{#1}}
\newcommand{\@defs@tens}[1]{\mathbf{#1}}

\DeclareMathOperator{\tr}{tr \!}

\newcommand{\Norm}[3]{\big|\!\big| #1 \big|\!\big|_{#2}^{#3}}

\newcommand{\de}{\,\mathrm{d}}
\newcommand{\ptl}{\partial}

\newcommand{\IR}{\mathbb{R}}

\newcommand{\Ctensor}{\mathds{C}}

\newcommand{\vect}[1]{\@defs@vec{#1}{}}
\newcommand{\tens}[1]{\@defs@tens{#1}{}}

\newcommand{\trp}{\textsf{T}}

\newcommand{\grad}{\boldsymbol{\nabla}}

\newcommand{\OmegaL}{\Omega_0}
\newcommand{\OmegaLP}{\Omega_{0,\text{C}}}

\newcommand{\SEI}{\text{S}}

\newcommand{\halb}{\frac{1}{2}}

\newcommand{\dualreduction}{\!:\!}

\newcommand{\ch}{\text{ch}}

\newcommand{\el}{\text{el}}
\newcommand{\cmax}{c_{\text{max}}}

\newcommand{\pl}{\text{pl}}
\newcommand{\gradL}{\grad_0}
\newcommand{\divgL}{\grad_0 \!\cdot\!}
\newcommand{\something}{\boldsymbol{\cdot}}
\newcommand{\minus}{\scalebox{0.75}[1.0]{$-$}}

\newcommand{\ext}{\text{ext}}
\newcommand{\devi}{\text{dev}}

\newcommand{\q}{\quad}

\newcommand{\conb}{\OB{c}}

\newcommand{\OB}[1]{\mkern 1.5mu\overline{\mkern-1.5mu#1\mkern-1.5mu}\mkern
  1.5mu}

\newcommand{\pc}{\textbf{CE}}
\newcommand{\plr}{\textbf{LR}}
\newcommand{\pul}{\textbf{UL}}

\newcommand{\ptcl}{\text{C}}
\newcommand{\sei}{\text{S}}

\makeatother

\title{\raggedright Elliptical Silicon Nanowire
  Covered by the SEI
  in a 2D
  Chemo-Mechanical Simulation
}

\author{
\begin{minipage}{0.98\textwidth}
  Raphael Schoof,*\textsuperscript{,+,[a]}
  Lukas Köbbing,*\textsuperscript{,+,[b,c]}
  Arnulf Latz,\textsuperscript{[b,c,d]}
  Birger Horstmann,\textsuperscript{[b,c,d]}
  Willy Dörfler\textsuperscript{[a]}
\end{minipage}
}

\newcommand{\affiliation}{
\begin{itemize}

\item[{[a]}] R. Schoof, Prof. Dr. W. Dörfler\\
Karlsruhe Institute of Technology~(KIT),
Institute for Applied and Numerical Mathematics,
Englerstraße~2, 76131~Karlsruhe, Germany\\
E-mail: raphael.schoof@kit.edu

\item[{[b]}] L. Köbbing, Prof. Dr. A. Latz, Prof. Dr. B. Horstmann\\
German Aerospace Center (DLR), Wilhelm-Runge-Straße 10, 89081 Ulm, Germany\\
E-mail: lukas.koebbing@dlr.de

\item[{[c]}] L. Köbbing, Prof. Dr. A. Latz, Prof. Dr. B. Horstmann\\
Helmholtz Institute Ulm (HIU), Helmholtzstraße 11, 89081 Ulm, Germany

\item[{[d]}] Prof. Dr. A. Latz, Prof. Dr. B. Horstmann\\
Faculty of Natural Sciences, Ulm University, Albert-Einstein-Allee 47, 89081
Ulm, Germany

\item[{[\texttt{+}]}] These authors contributed equally.
\end{itemize}
}

\renewcommand{\dedication}{
}

\renewcommand{\abstract}{
  Understanding the mechanical interplay between silicon anodes and
  their surrounding
  solid-electrolyte interphase~(SEI)
  is essential to improve the next generation
  of lithium-ion batteries.
  We model and simulate a 2D elliptical silicon nanowire with SEI via a
  thermodynamically consistent chemo-mechanical continuum ansatz using
  a higher order finite element method in combination with a variable-step,
  variable-order time integration scheme.
  Considering a soft viscoplastic SEI for three half cycles,
  we see at the minor half-axis the largest stress magnitude at the
  silicon nanowire surface, leading to a concentration anomaly.
  This anomaly is caused by the shape of the nanowire itself and not by the
  SEI.
  Also for the tangential stress of the SEI,
  the largest stress magnitudes are at this point,
  which can lead to SEI fracture.
  However, for a stiff SEI,
  the largest stress magnitude inside the nanowire occurs at the major half-axis, causing a reduced concentration distribution in this area.
  The largest tangential stress of the SEI is still at the minor
  half-axis.
  In total, we demonstrate the importance of considering the mechanics of the
  anode and SEI in silicon anode simulations and encourage further numerical
  and model improvements.
}

\newcommand{\keywords}{
	Silicon-SEI mechanics \textbullet\
	Stress distribution \textbullet\
	Lithiation characteristics \textbullet\
	Chemo-mechanical simulation \textbullet\
	2D elliptical nanowire
}

\begin{document}

\twocolumn[\vspace{-1.5cm}\maketitle\vspace{-1cm}
	\textit{\dedication}\vspace{0.4cm}
]
\small{\begin{shaded}
		\noindent\abstract
	\end{shaded}
}

\begin{figure} [!b]
\begin{minipage}[t]{\columnwidth}{\rule{\columnwidth}{1pt}\footnotesize{\textsf{\affiliation}}}\end{minipage}
\end{figure}





\section{Introduction}
\label{sec:introduction}

Silicon anodes can present the next vital step towards
improved
lithium-ion batteries with higher capacity
\cite{obrovac2014alloy,sun2022recent,zuo2017silicon,feng2018silicon-based,li2023si-based}.
 Nevertheless, the significant ability for lithiation causes massive
volume changes during cycling, hindering the commercialization of pure
silicon anodes \cite{beaulieu2001colossal,kim2023issues}. The
substantial deformations lead to mechanical instabilities of anode
particles larger than $\SI{150}{nm}$ and cause particle fracture and
pulverization \cite{liu2012size-dependent,wetjen2018morphological}.
Consequently, hopes are pinned on nanostructured silicon anodes
\cite{su2014silicon-based,yang2020silicon-nanoparticle-based,xu2023analysis} and
silicon nanowires in particular
\cite{chan2008high-performance,yang2020review,kilchert2024silicon}.

Due to electrolyte instability in contact with anode particles, the
solid-electrolyte interphase (SEI) forms on silicon anodes, reasonably
passivating the electrolyte from further decomposition
\cite{horstmann2019review,zhang2021interplay,adenusi2023lithium,wang2018review}.
However, the SEI continues to grow during storage and battery operation via
electron transport from the anode towards
the electrolyte
\cite{kolzenberg2020solid-electrolyte,kobbing2023growth,yao2023nucleation}.
On silicon anodes,
the SEI and its mechanical behavior merit special attention as the massive
volume changes of the anode challenge the stability of the SEI
\cite{zhang2019designing,kolzenberg2022chemo-mechanical}. Nonetheless, the
inner SEI is reported to stay intact during cycling \cite{guo2020failure}.
Thus, it is important to consider the stress generated inside the SEI and its
implication for silicon anodes in simulations of the silicon-SEI system
\cite{kobbing2024voltage,kobbing2024slow,wycisk2024challenges}.

Our previous works discussed the silicon-SEI system with spherical symmetry
\cite{kolzenberg2022chemo-mechanical,kobbing2024voltage,kobbing2024slow,schoof2024comparison}.
 Additionally, we performed 2D simulations of the nanowire only
\cite{castelli2021efficient, schoof2024modeling, schoof2022parallelization}
and restricted expansion by
a rigid obstacle \cite{schoof2023simulation}. As literature reports the
importance of non-symmetric geometries on the mechanical properties during
cycling \cite{tanaka2018role} in contrast to a
spherical setup~\cite{ding2024investigating}, we investigate the mechanics of an elliptical
silicon nanowire covered by SEI in this manuscript.
Therefore,
we straightforwardly adapt our 1D radial symmetric setup for the chemical and
elastic silicon core as well as the elastic and viscoplastic SEI shell to the
2D elliptical nanowire.
Our variable-step, variable-order time integration scheme is combined with a
higher order finite element method.
In total, we simulate three half cycles, meaning a first lithiation is
followed by delithiation and a second lithiation.
We provide extensive investigations of the mechanical characteristics and
concentration distribution for the coupled silicon-SEI structure.

The remaining of this manuscript is structured as follows:
in~\cref{sec:theory}, we present the key details of our continuum modeling,
followed by a brief summary of our numerical procedure in~\cref{sec:numerics}.
The focus of this work is~\cref{sec:results},
in which we present our extensive numerical results and discussions.
We conclude with a summary and a short outlook in~\cref{sec:conclusion}.




\section{Theory}
\label{sec:theory}
\begin{figure}[t]
  \centering
  \includegraphics[width = 0.95\textwidth,
  page=1]{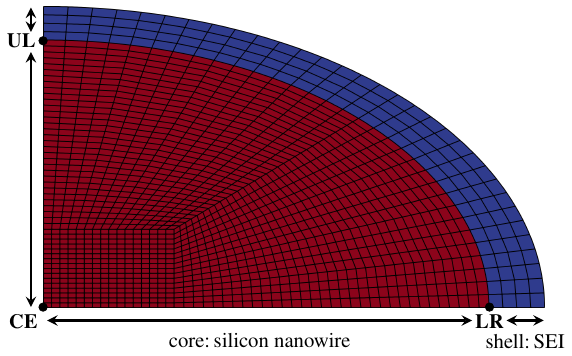}
  \caption{
    Sketch of the silicon nanowire core (red)
    covered by the SEI shell (blue)
    with the underlying time-constant computational grid
    and the points~$\plr$ (lower right), $\pul$ (upper left) and $\pc$ (center)
    for further investigations.
  }
  \label{fig:sketch}
\end{figure}

We follow Ref.~\cite{schoof2024comparison} and briefly recap our
chemo-mechanically
coupled model for the \emph{silicon-SEI}
approach.
This ansatz is
based on the thermodynamically consistent theory
by Refs.~\cite{schoof2024comparison, kobbing2024voltage, schoof2024modeling,
schoof2024residual,
kolzenberg2022chemo-mechanical, castelli2021efficient, schammer2021theory}.

We use a purely elastic (Lagrangian) logarithmic Hencky	strain for the
finite deformation model of the electrode nano\-wire, whereas we
apply a
viscoplastic approach for the SEI.
For a purely elastic particle without SEI, the typically used
Lagrangian
strain
or Green--St-Venant strain leads to similar results compared to the Hencky
strain~\cite{schoof2024modeling, zhang2018sodium}.
The deformation~$\Phi$
relates the reference (Lagrangian)
configuration~$\OmegaL \subset \IR^3$
to the current (Eulerian)
configuration~$\Omega$.
A silicon core subdomain and a SEI shell subdomain are
identified in each frame,
indicated with the
subscript $\ptcl$ and $\sei$, respectively.
In this work,
we consider a quarter section of an elliptical nanowire,
resulting from symmetry assumptions along both half-axes as well as
free expansion and vanishing stresses in the third direction.
At the half-axes, we impose a non-displacement condition in tangential
direction.
A sketch of the considered domain with the underlying computational grid is depicted
in~\cref{fig:sketch}.

\textbf{Finite Deformation.}
The deformation gradient
$
  \tens{F}
  =
  \tens{Id} + \gradL \vect{u}
$
with the identity
tensor~$\tens{Id}$
and the displacement
vector~$\vect{u}$
can be split up multiplicatively into three parts:
$\tens{F} = \tens{F}_\ch\tens{F}_\el\tens{F}_\pl$,
the chemical, elastic, and plastic deformation,
respectively,
compare
Sect.~10.4
in Ref.~\cite{bertram2021elasticity},
Sect.~8.2.2
in Ref.~\cite{lubliner2006plasticity}
and
Ref.~\cite{di-leo2015diffusion-deformation}.

In the silicon core domain,
we consider	only reversible
deformations~$\tens{F} = \tens{F}_\ch\tens{F}_\el = \tens{F}_\text{rev}$.
The elastic part results from mechanical stresses
and the chemical part
from changes of the lithium concentration during lithiation and delithiation
as~$\tens{F}_\ch = \lambda_\ch \tens{Id}
= \sqrt[3]{1+ v_\text{pmv} \cmax \conb} \, \tens{Id}$
with the partial molar
volume~$v_\text{pmv}$
of lithium inside silicon,
 the normalized lithium concentration~$\conb = c / \cmax \in [0, 1]$
of the lithium concentration~$c \in [0, \cmax]$
with respect to the maximal concentration~$\cmax$
in the
reference configuration.
In the SEI domain, no chemical deformation occurs:
$\tens{F} = \tens{F}_\el \tens{F}_\pl$.
We omit the index~$\ptcl$
or~$\SEI$
for reasons of better readability
if it is clear from the context which part is referred to.

\textbf{Free Energy.}
We consider all model equations in the reference configuration at constant
temperature and state the Helmholtz free
energy~$\psi(\conb, \gradL \vect{u}, \tens{F}_\pl)
=
\psi_\ch(\conb)
+
\psi_\el(\conb, \gradL \vect{u},
\linebreak
\tens{F}_\pl)$
resulting
in~$\psi(\conb, \gradL \vect{u}_\ptcl)
=
\psi_\ch(\conb)
+
\psi_{\el}(\conb, \gradL \vect{u}_\ptcl)$
and $\psi(\gradL \vect{u}_\SEI, \tens{F}_\pl)
=
\psi_{\el}(\gradL \vect{u}_\SEI, \tens{F}_\pl)$
for the respective silicon core and SEI shell domain.
With the mass density~$\rho_0$
of silicon in the reference configuration
the chemical and elastic free energy densities can be defined as
\begin{align*}
  \rho_0\psi_\ch(\conb)
  =
  \minus
  \cmax
  \int_0^{\conb} F \, U_\text{OCV}(z) \de z
\end{align*}
with an experimental open-circuit
voltage~(OCV)
curve
\linebreak
$U_\text{OCV}$~\cite{kolzenberg2022chemo-mechanical, kobbing2024voltage,
  schoof2024modeling, schoof2023simulation}
and
the Faraday
constant~$F$
as well as
\begin{align*}
  \rho_0 \psi_\el (\conb, \gradL \vect{u}, \tens{F}_\pl)
  =
  \halb \tens{E}_\el(\conb, \gradL \vect{u}, \tens{F}_\pl) \dualreduction
  \Ctensor[\tens{E}_\el]
\end{align*}
with~$\tens{F}_\pl = \tens{Id}$
for silicon,
the elastic strain tensor~$\tens{E}_\el$,
and the constant, isotropic stiffness fourth-order
tensor~$\Ctensor$
as~$\Ctensor[\tens{E}_\el]
= \lambda \tr \left(\tens{E}_\el\right) \tens{Id}
+
2G \tens{E}_\el$.
Here,
$\lambda = 2G \nu / (1-2\nu)$
and~$G = E / \big( 2(1+\nu)\big)$
are the first and second Lam\'e
constants, respectively,
depending further on Young's modulus~$E$
and Poisson's ratio~$\nu$.
In~\cref{tab:parameters},
we give the parameters for silicon and SEI.
The (Lagrangian) logarithmic Hencky strain
tensor~$\tens{E}_\el$
is given as
$\tens{E}_\el
= \ln \! \left(\tens{U}_\el\right)
= \ln \! \left(\sqrt{\tens{C}_\el} \right)
= \sum_{\alpha = 1}^{3}
\ln \! \left(\sqrt{\eta_{\el, \alpha}}\right)
\vect{r}_{\el, \alpha} \otimes \vect{r}_{\el, \alpha}$
with the
eigenvalues~$\eta_{\el, \alpha}$
and
eigenvectors~$\vect{r}_{\el, \alpha}$
of~$\tens{U}_\el$.
The tensor~$\tens{U}_\el$ is the unique, symmetric and positive definite right
stretch part of the unique polar decomposition
of~$\tens{F}_\el =
\tens{R}_\el\tens{U}_\el$,
see Sect.~2.6
in Ref.~\cite{holzapfel2010nonlinear}.

\textbf{Chemistry.}
The lithium concentration changes during lithiation and delithiation inside the
reference silicon core domain~$\OmegaLP$
can be stated via a generalized diffusivity
equation~\cite{anand2012cahn-hilliard-type, di-leo2015chemo-mechanics,
  kolzenberg2022chemo-mechanical}
\begin{align}
  \ptl_t c
  =
  \minus \divgL \vect{N}.
  \label{eq:conti}
\end{align}
The lithium
flux~$\vect{N}
= \minus D \left(\ptl_c \mu \right)^{-1} \gradL \mu$
with
the diffusion
coefficient~$D$ for lithium in silicon
is applied for an isotropic case.
The chemical potential~$\mu$
can be derived as the partial derivative of the free energy density with respect
to the concentration~$c$~\cite{kolzenberg2022chemo-mechanical, kobbing2024voltage,
schoof2024modeling, schoof2023simulation, schoof2024residual}
\begin{align}
  \mu
  =
  \ptl_c (\rho_0 \psi)
  &
  =
  \mu_\ch + \mu_\el
  \nonumber
  \\
  &
  =
  \minus F \, U_\text{OCV}
  -
  \frac{v_\text{pmv}}{3 \lambda_\ch^3} \tr \left(\Ctensor[\tens{E}_\el]\right).
  \label{eq:chemical_potential}
\end{align}
Therefore, the total lithium flux~$\vect{N} = \vect{N}_\ch + \vect{N}_\el$
can be divided into the lithium concentration-driven diffusive
flux component $\vect{N}_\ch
=
\minus D \left(\ptl_c \mu \right)^{-1} \gradL \mu_\ch$
and the stress-driven convective flux component $\vect{N}_\el
=
\minus D \left(\ptl_c \mu \right)^{-1} \gradL \mu_\el$,
respectively.
A uniform and constant external flux~$N_\ext$ in the Lagrangian domain
with either positive or negative
sign (for lithiation or delithiation, respectively) is applied at the surface
of the silicon core.
This external flux is measured with regard to the charging rate
(\si{C}-rate) connecting the state of charge (SOC) to the simulation time via
the external lithium flux
and the initial concentration
$\text{SOC}(t) = \conb_0 + N_\ext t$.
Further information about the SOC, the  \si{C}-rate and $N_\ext$ can be found
in Refs.~\cite{schoof2024modeling, kolzenberg2022chemo-mechanical,
  castelli2021efficient}
and the references cited therein.

\textbf{Elastic and Inelastic Deformation.}
We solve the momentum balance
equation~\cite{kolzenberg2022chemo-mechanical, castelli2021efficient,
schoof2024modeling, schoof2023simulation} in the silicon core domain and the
SEI shell domain
\begin{align}
  \vect{0}
  =
  \divgL \tens{P}_\ptcl(\conb, \gradL \vect{u}_\ptcl),
  \qquad
  \vect{0}
  =
  \divgL \tens{P}_\SEI(\gradL \vect{u}_\SEI, \tens{F}_\pl)
  \label{eq:momentum_balance}
\end{align}
for the respective deformation.
The first Piola--Kirchhoff
tensor~$\tens{P}$ is
thermodynamically consistently derived
as~
\begin{align*}
\tens{P}
= 2 \tens{F}\ptl_\tens{C}(\rho_0 \psi)
= \tens{F}
\left(\tens{F}_\el^\trp \tens{F}_\el^{\vphantom{\trp}}\right)^{-1}
\left(\tens{F}_\pl^{-1} \right)^{\trp}
\tens{F}_\pl^{-1}
\Ctensor[\tens{E}_{\el}],
\end{align*}
see Refs.~\cite{kolzenberg2022chemo-mechanical,
schoof2024modeling,
  schoof2023simulation, kobbing2024voltage}.
With the first Piola--Kirchhoff tensor~$\tens{P}$
we state the related symmetric Cauchy
stress~$\boldsymbol{\sigma}$
in the
current configuration
as~$\boldsymbol{\sigma}
= \tens{P} \tens{F}^\trp /
\det \left(\tens{F}\right)$,
see Sect.~3.1
in Ref.~\cite{holzapfel2010nonlinear}.

In this work, we rely on the rate-dependent plastic
approach~\cite{schoof2024comparison, schoof2024modeling}.
Therefore, we introduce
the scalar yield
\linebreak stress~$\sigma_\text{Y}$
and the
evolution equation of the scalar accumulated equivalent inelastic
strain~${\varepsilon}_\pl^\text{eq} \geq 0$
as
\begin{subequations}
  \label{eq:viscoplatic_epsplver}
  \begin{empheq}[
    left={
      \dot{\varepsilon}_\pl^\text{eq} =
      \empheqlbrace}
    ]{alignat=3}
    &0,
    \nonumber
    && \Norm{\tens{M}^{\devi}}{}{} \leq \sigma_\text{Y},
    \\
    &
    \dot{\varepsilon}_0
    \Bigg(
    \frac{\Norm{\tens{M}^{\devi}}{}{} - \sigma_\text{Y}}
    {\sigma_{\text{Y}^{*}}}\Bigg)^\beta,
    \q
    &&
    \nonumber
    \Norm{\tens{M}^{\devi}}{}{} > \sigma_\text{Y},
  \end{empheq}
\end{subequations}
which replace the typical Karush--Kuhn-Tucker (KKT) conditions for the plastic
approach,
compare~Sect.~1.7
in Ref.~\cite{simo1998computational}
and Refs.~\cite{di-leo2015diffusion-deformation,
  schoof2024modeling, schoof2024comparison}.
The deviatoric Mandel
stress~$\tens{M}^{\devi}
= \tens{M} - 1/3 \tr\left( \tens{M} \right) \tens{Id}$
is computed via the Mandel stress~$\tens{M}
=
\linebreak
\ptl_{\tens{E}_\el} (\rho_{0} \psi_{\el})
= \Ctensor_\SEI[\tens{E}_{\el}]$
in the SEI domain.
The remaining values
are the positive-valued stress-dimensioned
constant~$\sigma_{\text{Y}^{*}}$,
the reference tensile
stress~$\dot{\varepsilon}_0$,
and the measure of the strain rate sensitivity of the
material~$\beta$ which are given in~\cref{tab:parameters}.
Furthermore, we rescale the yield stress with the
factor~$\sqrt{2/3}$ due to consistency with the one dimensional tensile
test,
see Sect.~2.3.1
in Ref.~\cite{simo1998computational}.
Finally, we use a projector formulation
to map the stresses onto the set of admissible stresses,
stated for our viscoplastic approach in Ref.~\cite{schoof2024modeling}.
This procedure is also known as static
condensation~\cite{wilson1974static, di-pietro2015hybrid}.
Therefore, $\tens{F}_\pl$
and~$\varepsilon_{\pl}^\text{eq}$
are applied as internal variables.
This procedure has the advantage that the nonlinear system of partial
differential
equations does not need to be extended by the plastic part of the deformation
gradient,
in contrast to Refs.~\cite{kobbing2024voltage, kolzenberg2022chemo-mechanical}.




\section{Numerical Approach}
\label{sec:numerics}

Again, we follow Ref.~\cite{schoof2024comparison}
and state only the most important details.
All in all,
after non-dimensionalization and omitting the accentuation for the
non-dimensionalization,
we solve
for given~$\tens{F}_\pl$
and~$\varepsilon_{\pl}^\text{eq}$
the continuity equation
in~\cref{eq:conti},
the chemical potential equation
in~\cref{eq:chemical_potential},
and the momentum balance equations
in~\cref{eq:momentum_balance}.
As a result, we obtain the concentration~$c$, the chemical potential~$\mu$, and the silicon core displacement~$\vect{u}_\ptcl$
as well as the SEI shell displacement~$\vect{u}_\sei$.
Therefore, we imply boundary conditions at the interface between the
silicon core and the SEI shell domain:
$\vect{u}_\ptcl = \vect{u}_\SEI$
and~$\tens{P}_\ptcl \cdot \vect{n} = \tens{P}_\SEI \cdot \vect{n}$
with the normal vector~$\vect{n} = \vect{n}_\ptcl$.
At the outer boundary of the SEI, we have no stresses
meaning~$\tens{P}_\SEI \cdot \vect{n}_\SEI = \vect{0}$.
Furthermore, we impose initial conditions~$c(0, \something)=c_0$,
$\tens{F}_{\pl}(0, \something) = \tens{Id}$
and~$\varepsilon_{\pl}^\text{eq}(0, \something) = 0$.

For the numerical solution of the nonlinear system of partial differential
equations,
we choose an admissible mesh for the computational domain,
use the isoparametric Lagrangian finite element
method,
see Chapt.~III~§2
in Ref.~\cite{braess2007finite},
derive a weak formulation
and a spatial and temporal
discretization~\cite{schoof2024comparison}.
For the spatial discretization,
we apply a fourth order finite element approach
using a uniform and time-constant mesh in the reference configuration,
displayed
in~\mbox{\cref{fig:sketch}}.
Note that the original set of equations is derived in 3D,
however, all
equations are also mathematically valid in 2D~\cite{castelli2021efficient,
schoof2024modeling}.
The temporal discretization is realized with a variable-step, variable-order
time integration scheme using the numerical differential formulation~(NDF)
of linear multistep
methods~\cite{reichelt1997matlab, shampine1997matlab,
  shampine1999solving}.
The temporal discretization of the internal variables are treated with an
implicit exponential map.
For a detailed procedure of the temporal integration
for~$\tens{F}_{\pl}$
and~$\varepsilon_{\pl}^\text{eq}$,
we refer to Ref.~\cite{schoof2024modeling}.
In each time step, the nonlinear system is solved using the Newton--Raphson
method and the adaptive scheme for the time
presented as Algorithm 1 in Ref.~\cite{castelli2021efficient}.

We start with the constant initial
concentration~$\conb_0 = 0.02$
and~$\mu_0 = \ptl_c \rho\psi_\ch (\conb_0)$.
The initial time step size is~$\SI{e-8}{h}$,
the maximal time step size~$\SI{e-3}{h}$
and temporal relative and absolute tolerances
$\num{2e-4}$ and $\num{2e-7}$,
respectively.
The grid has around $\num{87e3}$ degrees of freedom.
Additional zero-displacement boundary conditions are applied on the major
half-axis
with~$u_y = 0$
and on the minor half-axis
with~$u_x = 0$.
The Newton update is computed with an LU-decomposition from the UMFPACK
package~\cite[Version~5.7.8]{davis2004algorithm}
and shared memory with OpenMP Version 4.5 is enabled for assembling the Newton
method.
Our implementation is based on the open-source finite element library
\textit{deal.II}~\cite{arndt2023deal}.
All simulations are performed on a single node at the
Bw\-Uni\-Cluster2.0 with GCC 12.1~\cite{wiki-bw-hpc-team2024bwunicluster}.




\section{Results and Discussion}
\label{sec:results}

Due to the importance of mechanics and the silicon anode geometry, we
investigate the chemo-mechanical coupling of an elliptical silicon nanowire
covered by the SEI in a 2D setup. We discuss the stresses occurring inside the
silicon anode and the SEI in comparison to a symmetric nanowire. Additionally,
we examine the lithium concentration distribution and gradients during
lithiation and delithiation influenced by mechanics. To assess the impact of
the SEI, we compare the chemo-mechanical results for a silicon anode covered by
a soft and a stiff SEI layer.

During cycling, the lithium concentration inside the silicon nanowire changes.
An increase in the lithium concentration results in a chemical expansion of the
anode, while a decrease leads to a shrinkage. Inhomogeneous lithium
distribution inside the silicon implies inhomogeneous volume changes that have
to be accommodated by mechanical deformations. These mechanical strains inside
the lithiated silicon generate stresses.
While the silicon can deform chemically and elastically, the SEI features elastic
and viscoplastic material behavior.
During cycling, the SEI layer has to
adjust to the volume changes of the silicon anode. As the SEI
can only deform mechanically, expansion and shrinkage of the silicon anode lead
to significant mechanical strains, creating stresses inside the SEI as well.
The stresses inside silicon and SEI are coupled due to the interface condition
of equal stress in
normal direction.

\subsection{Silicon Nanowire with Soft SEI}

\begin{figure*}[t]
	\centering
	\includegraphics[width = 0.95\textwidth,
	page=1]{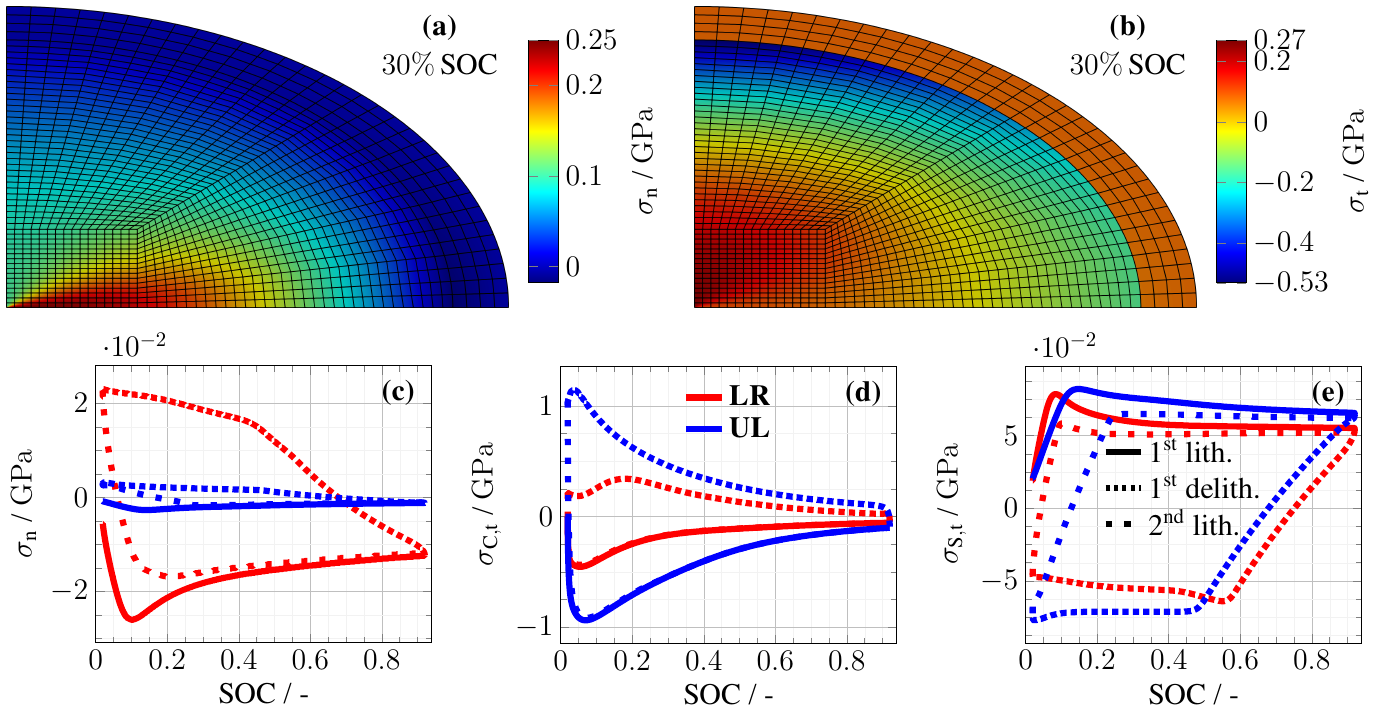}
	\caption{Cauchy stresses for the elliptical silicon nanowire with a soft SEI. Distribution of (a) normal and (b) tangential Cauchy stress inside the silicon core and the SEI shell during lithiation at $30\%$ SOC. Evolution of Cauchy stress at the points $\plr$ and $\pul$ during three half cycles for (c) normal, (d) tangential core, and (e) tangential shell stress.}
	\label{fig:sigma_soft_sei}
\end{figure*}

First, we investigate the behavior of an elliptical silicon nanowire covered by
a soft SEI layer.
The ratio of the minor to the major half-axis is 0.6:1.
The SEI thickness is one eighth of the silicon core length.
We simulate three half cycles ($1^\mathrm{st}$ lithiation,
$1^\mathrm{st}$ delithiation, and $2^\mathrm{nd}$ lithiation) with a rate of
$1\,\si{C}$.
The simulation parameters are stated in \cref{tab:parameters}. We begin our
discussion with the mechanics inside the silicon anode and the SEI layer and
continue with the examination of the lithium concentration distribution.
Especially, we consider the quantities of interest at the lower right of the major half-axis (point~$\plr$), at the upper left of the
minor half-axis (point~$\pul$), and at the center of the silicon (point~$\pc$), respectively. The interfacial points ($\plr$ and $\pul$) and the central point ($\pc$) are
illustrated in~\cref{fig:sketch}.

Concerning the mechanics inside the silicon anode, lithiation from the outside
leads to concentration gradients and inhomogeneous volume changes. The volume
mismatch generates compressive stress at the outer boundary of the nano\-wire
and tensile stress at the center. During delithiation, lithium flux out of the
anode leads to tensile stress at the outer boundary and compressive stress at
the center.
To investigate the mechanics in detail, we illustrate the simulated stresses for the elliptical silicon nanowire covered by a soft SEI layer in \cref{fig:sigma_soft_sei}. We depict the stress distribution during lithiation at $30\%$ SOC for the normal component $\sigma_\mathrm{n}$ in \cref{fig:sigma_soft_sei}(a) and the tangential component $\sigma_\mathrm{t}$ in \cref{fig:sigma_soft_sei}(b). Both stress distributions reveal the general trend of tensile stress at the center and compressive stress at the outer boundary of the silicon nanowire. The largest compressive stresses appear
in normal direction at the end of the major half-axis at point $\plr$ and in tangential direction at the end of the minor half-axis at point $\pul$.
The largest tensile stresses appear
in normal direction along the major half-axis and in tangential direction along
the minor half-axis, each close to the center. Therefore, possible plasticity
\cite{schoof2024modeling}
and fracture might occur along the minor half-axis.

We depict the time evolution of the stress inside silicon during three half cycles for the
normal component in \cref{fig:sigma_soft_sei}(c) and for the tangential
component in \cref{fig:sigma_soft_sei}(d). The evolution of the stress
components during the first lithiation reveals permanent compressive stress at
points $\pul$ and $\plr$, with significantly larger stress magnitudes for the tangential component. The normal stress in \cref{fig:sigma_soft_sei}(c)
always shows the largest magnitude at the major half-axis at point $\plr$, and
the tangential stress in \cref{fig:sigma_soft_sei}(d) shows the largest magnitude at the minor half-axis at
point $\pul$. During the subsequent delithiation, tensile stresses arise at the
outer boundary, showing the largest magnitudes at the same points as before.
The normal stress during the second lithiation
in~\cref{fig:sigma_soft_sei}(c)
deviates from the first lithiation in the beginning due to a different initial
state, but the stresses continuously approach the ones during the first
lithiation. The tangential stress during the second lithiation in~\cref{fig:sigma_soft_sei}(d) coincides with the first lithiation.
The largest stress magnitudes during cycling occur at low SOC in
particular in tangential direction at the end of the minor half-axis at point $\pul$. This supports our previous finding that the
elliptical silicon nanowire might be prone to plasticity
\cite{schoof2024modeling} and fracture in this regime.

To highlight the influence of the elliptical geometry, we compare our
simulation results to the case of a symmetric silicon nanowire with the same
capacity in \cref{fig:sigma_symmetric}. Due to the symmetry, the stresses at
points $\pul$ and $\plr$ are equal. The stress magnitudes of the normal and
tangential stress components at the outer boundary in the symmetric case are
always in between the stress values at point $\pul$ and $\plr$ for the
elliptical case. Therefore, the largest stresses reached during cycling in the
symmetric case stay smaller than the ones for the elliptical case.
Consequently, the symmetric silicon nanowires are mechanically more stable
than elliptical silicon nanowires with the same capacity.

Next, we discuss the mechanics of the SEI shell during cycling. The Cauchy
stress in normal direction in silicon and SEI is coupled at the interface.
At the outer
boundary of the SEI, the stress in normal direction vanishes. Therefore, we
focus on the description of the tangential component of the SEI stress depicted
in \cref{fig:sigma_soft_sei}(e). During lithiation, the volume expansion of the
nanowire leads to tensile tangential stress inside the SEI. The tangential
stress magnitude at the minor half-axis at point $\pul$ is slightly larger
compared to point $\plr$. This can be expected as the curvature of the SEI is
smallest at the end of the minor half-axis. Thus, the SEI might be prone
to fracture at point $\pul$. During delithiation, the tangential stress inside
the SEI is compressive and the maximum value is reached again at point $\pul$.
During the second lithiation, the size of the stress overshoot reduces and the
stress converges to that
one of the first lithiation. Compared to the
symmetric nanowire, we observe the same trend for the stresses inside the SEI
shell as for the silicon core. The stress magnitudes are in between the
stresses at points $\pul$ and $\plr$. Consequently, the maximum value is
smaller for the symmetric case, meaning a superior mechanical stability of
the SEI. To investigate the influence of viscosity, we vary the parameter for
the plastic strain rate $\dot{\varepsilon}_0$ in
\cref{fig:different_dot_eps_0_all}. A smaller value retards plastic flow,
leading to a larger stress overshoot and larger stress magnitudes in general.
Nevertheless, the shape of the stress profiles does not change significantly
upon variation of $\dot{\varepsilon}_0$.

\begin{figure*}[t]
	\centering
	\includegraphics[width = 0.95\textwidth,
	page=1]{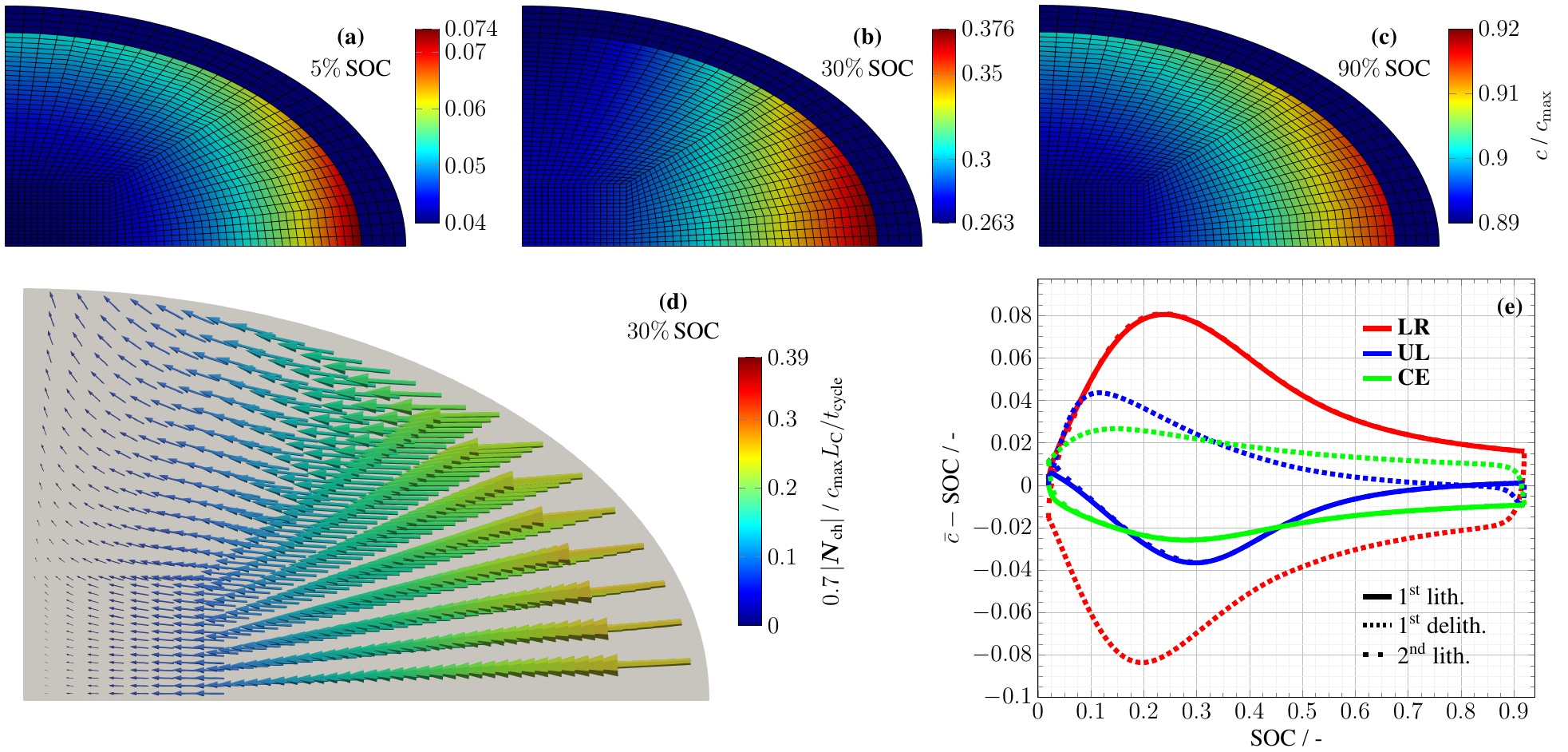}
	\caption{Lithium concentration for the elliptical silicon nanowire with a
	soft SEI. Distribution of the lithium concentration inside silicon during
	lithiation at (a) $5\%$, (b) $30\%$, and (c) $90\%$ SOC.
  (d)
  Concentration-driven diffusive lithium flux $\vect{N}_\ch$
  scaled with
  $0.7$.
  (e) Deviation of the lithium concentration from the mean at the points~$\plr$,
  $\pul$, and
  $\pc$ during three half cycles.}
  \label{fig:c_soft_sei}
\end{figure*}

After the mechanical description, we investigate the lithi\-um concentration
inside the elliptical silicon nanowire. During lithiation, the lithium flux
points from the outside into the interior of the silicon anode. Thus, we expect
that the concentration at the outer boundary of the anode exceeds the
concentration at the center, as shown for the symmetrical nanowire covered by
SEI in \cref{fig:c_symmetric}. During delithiation, we expect
a decreased lithium
concentration at the outer boundary compared to the center.
We show our simulation results for the lithium concentration in
\cref{fig:c_soft_sei}. The lithium distribution inside the elliptical silicon
nanowire during lithiation is illustrated in \cref{fig:c_soft_sei}(a) to (c)
for $5\%$, $30\%$, and $90\%$ SOC. As expected, the lithium concentration
increases in general from the outside. The lithium distribution reveals the
highest concentration at the end of the major half-axis at point $\plr$. At
this point, the elliptical geometry has the highest local surface-to-volume
ratio, resulting in faster lithium concentration increase,
compare Section~2.2.3~\cite{huttin2014phase-field}.
Contrary to our expectation, the lithium concentration at the end of the minor
half-axis at point $\pul$ is lower than the concentration at the center point
$\pc$ during lithiation at $30\%$ SOC.
For a better illustration of this
concentration anomaly,
we depict the concentration-driven diffusive lithium
flux~$\vect{N}_\ch$
during lithiation at $30\%$ SOC
in \cref{fig:c_soft_sei}(d),
indicating negatively scaled
concentration gradients.
Along the major half-axis,
lithium
diffusion points towards
the center of the ellipse, as expected. However, along the minor
half-axis, diffusion points towards
the outer boundary of the nanowire,
revealing the concentration depletion at point $\pul$.
While the arrows
indicating the direction of lithium diffusion partially point towards
the outer
boundary, the more pronounced stress-driven convective lithium flux
$\vect{N}_\el$
depicted in
\mbox{\cref{fig:stress_driven_lithium_flux}(a)} points towards the interior.
This ensures that the total flux $\vect{N}$ always points towards the interior
of the silicon core during lithiation.

To confirm the appearance of the concentration anomaly, we examine the
deviation of the lithium concentration from the mean during cycling in
\cref{fig:c_soft_sei}(e). As expected during
lithiation, the concentration at point $\plr$ is always larger, and at point
$\pc$ smaller than the mean concentration. During delithiation, this
concentration distribution is inverse. However, the concentration at point
$\pul$ is smaller than the mean concentration and even smaller than the
concentration in the center $\pc$ during lithiation in a wider SOC regime between $15\%$ and $45\%$ SOC. The
concentration anomaly at point
$\pul$ also appears during delithiation as concentration excess between $35\%$ and $5\%$ SOC and the second
lithiation again as depletion. Thus, the concentration anomaly at point $\pul$ is no simulation
artifact in a narrow SOC range during the first lithiation but significant and
robust during cycling.

The concentration anomaly also appears during slow cycling with $\mathrm{C}/20$
and inside an elliptical silicon nanowire without SEI as shown in
\cref{fig:c_particle_only_C20}. Thus, we exclude kinetic limitations or the
mechanical impact of the SEI on the silicon core as reasons for the
concentration anomaly. Instead, we attribute this effect to a mechanical origin
inside the elliptical silicon nanowire. During lithiation, the lithium
concentration increases the fastest at point $\plr$ at the end of the major
half-axis due to the highest surface-to-volume ratio. The significant increase
causes pronounced volume expansion, leading to compressive stress along the
outer boundary of the nanowire. This compressive stress is largest at point
$\pul$ at the end of the minor half-axis due to the smaller curvature at this
point. The substantial compressive stress affects the chemo-mechanical
potential and hinders further
lithium concentration increase
at point $\pul$. During delithiation,
the fastest decrease in concentration appears at point $\plr$,
generating tensile
stress, especially at point $\pul$. The substantial tensile stress impedes
lithium concentration decrease
at point $\pul$, generating a local concentration excess. Investigating a purely chemical 2D elliptical silicon nanowire without
mechanical coupling, such a concentration anomaly does not occur.
Thus, the concentration anomaly during cycling results from the
chemo-mechanical interplay inside the silicon nanowire significantly
influenced by the elliptical geometry.

\begin{figure*}[t]
	\centering
	\includegraphics[width = 0.95\textwidth,
	page=1]{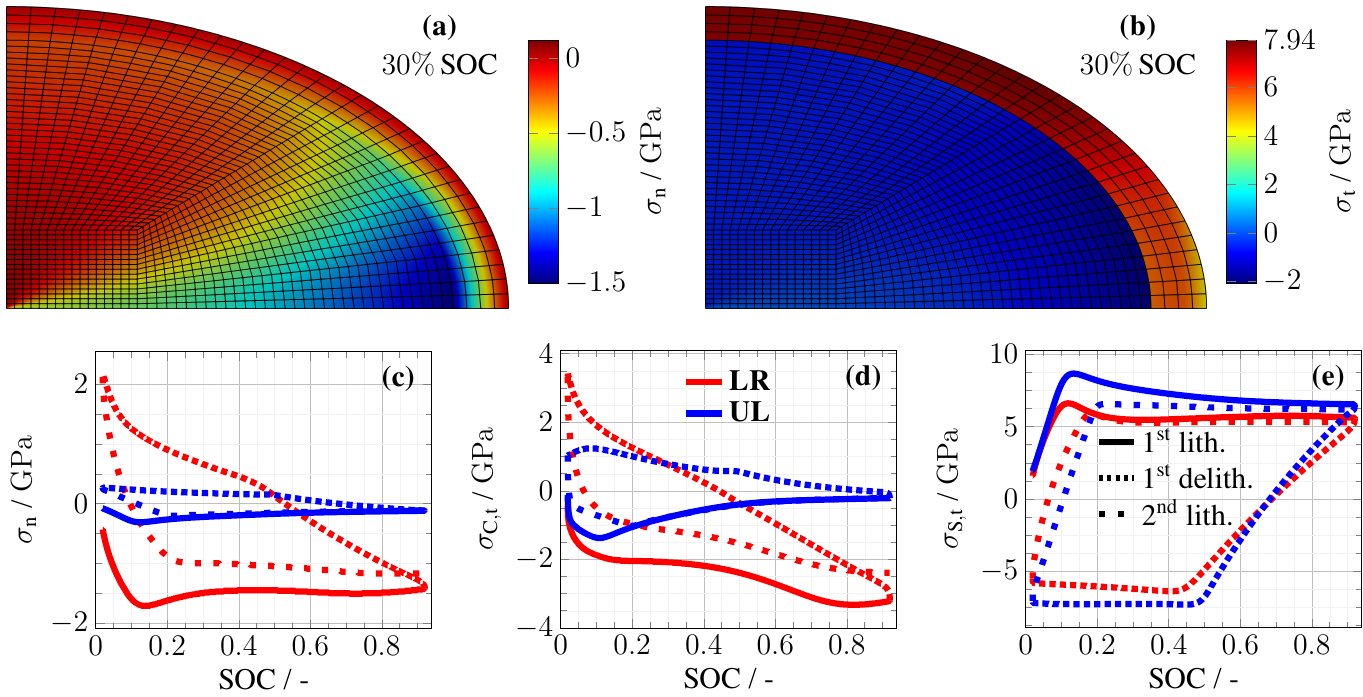}
	\caption{Cauchy stresses for the elliptical silicon nanowire with a stiff SEI. Distribution of (a) normal and (b) tangential Cauchy stress inside the silicon core and the SEI shell during lithiation at $30\%$ SOC. Evolution of Cauchy stress at the points $\plr$ and $\pul$ during three half cycles for (c) normal, (d) tangential core, and (e) tangential shell stress.}
	\label{fig:sigma_stiff_sei}
\end{figure*}

\subsection{Silicon Nanowire with Stiff SEI}

After discussing the soft SEI, we want to investigate the influence of a stiff
SEI layer on the mechanics and the lithiation behavior of an elliptical silicon
nanowire as discussed for a spherical nanoparticle in Ref.~\cite{kobbing2024voltage}.
Therefore, we increase the value of Young's modulus and the yield stress of the
SEI shell by a factor of 100 compared to the soft SEI, i.e. $E=\SI{90}{GPa}$
and $\sigma_\mathrm{Y}=\sigma_\mathrm{Y^*}=\SI{4.95}{GPa}$. The increase in the mechanical
parameters immediately evokes elevated stresses inside the SEI. We depict the
stress distribution in \cref{fig:sigma_stiff_sei}(a) for the normal component
and in \cref{fig:sigma_stiff_sei}(b) for the tangential component at $30\%$ SOC
during lithiation. Analog to the soft SEI scenario, the largest compressive
stress in the SEI in normal direction occurs at the major half-axis at
point $\plr$ and the largest tensile stress in tangential direction appears at
the minor half-axis at point $\pul$ due to the local curvature effects.
Therefore, possible cracking of the SEI might occur again at point $\pul$ due to the largest tangential stresses.

The time evolution during cycling of the normal stress in
\cref{fig:sigma_stiff_sei}(c) and the tangential stress in
\cref{fig:sigma_stiff_sei}(e) confirms this observation. During delithiation,
we observe the largest tensile stress in normal direction
inside the SEI at the major half-axis at
point~$\plr$ and the largest compressive stress in tangential direction at the
minor half-axis at point~$\pul$ accordingly. The stresses during the second
lithiation approach the stresses during the first lithiation but deviate due to
the viscoplastic behavior.
The comparison to the soft SEI case reveals a stress
increase inside the SEI for both components by approximately a factor of 100,
representing the increase in the mechanical parameters.

The stress inside the silicon nanowire is affected by the stiff SEI layer due
to the mechanical coupling of the silicon core and the SEI shell. We depict the
normal stress component inside silicon and SEI during lithiation at $30\%$ SOC
in \cref{fig:sigma_stiff_sei}(a). The illustration reveals that the normal
stresses at the interface are equal as imposed by the boundary condition. The
stress distribution shows significantly larger compressive stresses within the
whole silicon nanowire except a small region along the minor half-axis close to
the center, where tensile stresses appear. Compared to the soft SEI, the most
significant normal compressive stress occurs again at the end of the major
half-axis at point $\plr$ due to the largest curvature and pronounced impact of
the SEI. The tangential stress component inside silicon depicted in
\cref{fig:sigma_stiff_sei}(b) is indirectly affected by the different SEI
mechanics. The stress distribution reveals compressive stresses within the
whole silicon nanowire with the largest stress magnitude at the end of the
major half-axis at point $\plr$. This is in contrast to the case with the soft
SEI, where the largest compressive stress occurs at the end of the minor
half-axis at point $\pul$ and where tensile stresses occur in a larger region
around the center.

We depict the stress evolution within the silicon core during cycling in the
normal direction in \cref{fig:sigma_stiff_sei}(c) and in the tangential
direction in \cref{fig:sigma_stiff_sei}(d). As discussed for the SEI mechanics,
the normal stress inside silicon at the boundary is approximately 100 times
larger compared to the soft SEI scenario with a similar shape of the stress
profile. The tangential stress inside silicon is compressive during lithiation
and changes to tensile stress during delithiation analog to the soft SEI case.
However, the largest tangential stress magnitudes appear at point $\plr$ in
contrast to the scenario with the
soft SEI, where the largest tangential stress magnitude appears at point
$\pul$. Thus, the maximum stress magnitudes inside silicon occur at the same
point $\plr$ for the normal and tangential component due to the impact of the
stiff SEI shell, which is most significant at this point due to the largest
curvature. This effect emphasizes the importance of the mechanical interplay
between the silicon core and the SEI shell.

\begin{figure*}[t]
	\centering
	\includegraphics[width = 0.95\textwidth,
	page=1]{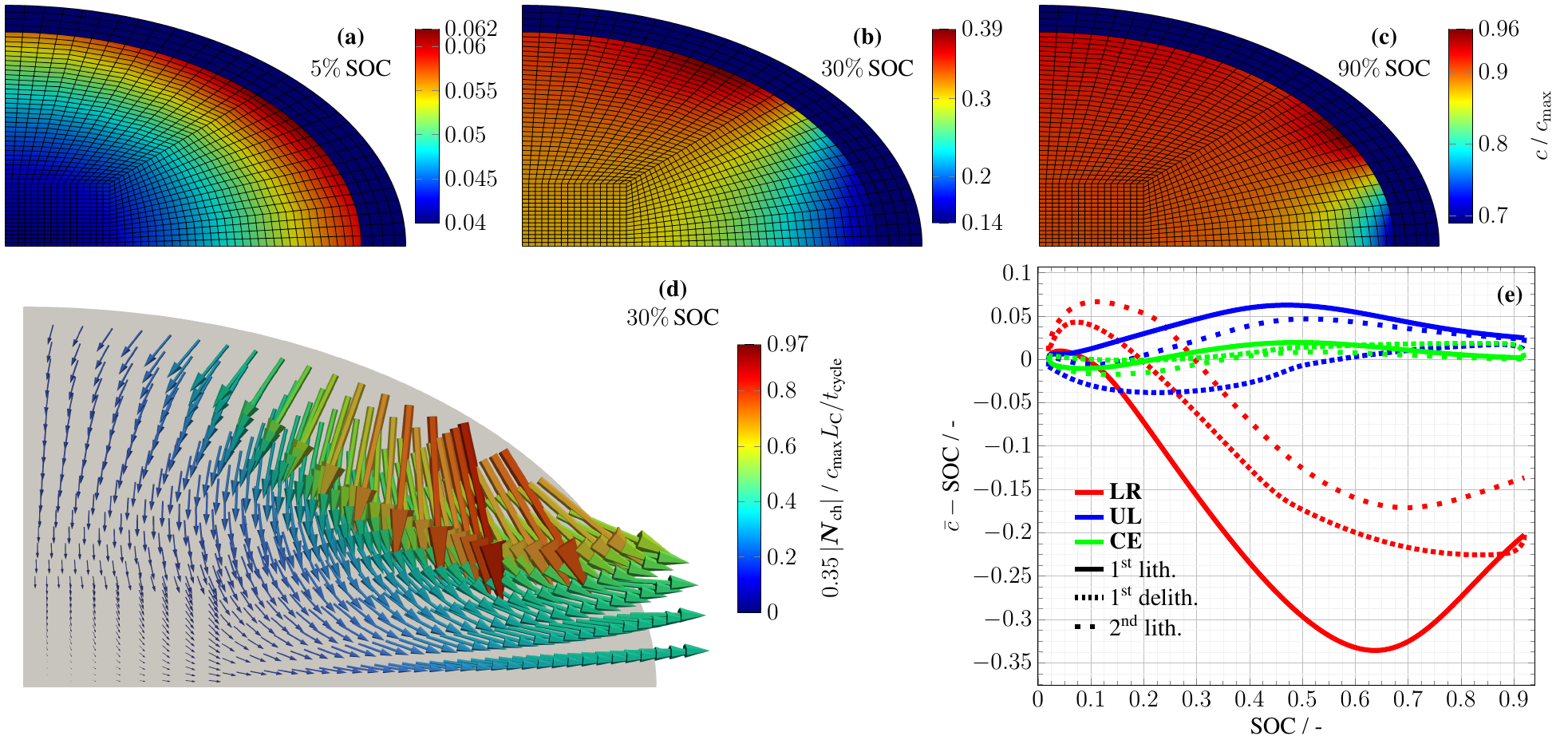}
	\caption{Lithium concentration for the elliptical silicon nanowire with a
	stiff SEI. Distribution of the lithium concentration inside silicon during
	lithiation at (a) $5\%$, (b) $30\%$, and (c) $90\%$ SOC.
    (d)
    Concentration-driven diffusive lithium flux $\vect{N}_\ch$
    scaled
    with~$0.35$.
    (e) Deviation of the lithium concentration from the mean at
    the points $\plr$,
    $\pul$, and $\pc$ during three half cycles.}
  \label{fig:c_stiff_sei}
\end{figure*}

Next, we discuss the influence of the stiff SEI mechanics on the lithiation
behavior of the silicon nanowire core. We depict the lithium concentration
distribution during lithiation in \cref{fig:c_stiff_sei} at (a) $5\%$, (b)
$30\%$, and (c) $90\%$ SOC. The illustration reveals that upon the start of the
lithiation, lithium concentration increases at the outer boundary of the
silicon core, proceeding gradually towards
the center as expected.
Nevertheless,
this trend is broken during further lithiation, and a concentration anomaly
occurs at the end of the major half-axis at point $\plr$.
The concentration-driven diffusive lithium flux $\vect{N}_\ch$ during
lithiation at $30\%$ SOC depicted as arrows in \cref{fig:c_stiff_sei}(d)
indicates the negatively scaled concentration gradient and confirms the
concentration depletion at point $\plr$.
This is in contrast to the
concentration anomaly found for the soft SEI case at the end of the minor
half-axis at point $\pul$, where no anomaly occurs for the stiff SEI case.
Instead, the anomaly appears at the point with the largest curvature and the
most significant stress magnitude generated by the stiff SEI.
Again, the stress-driven convective lithium flux $\vect{N}_\el$ depicted
in \cref{fig:stress_driven_lithium_flux}(b) guarantees that the total lithium
flux $\vect{N}$ always points towards the interior of the silicon core during
lithiation.
This confirms the
importance of the chemo-mechanical interplay and the severe influence of the
stiff SEI shell on the lithiation behavior of the silicon nanowire core. The
stiff SEI shell acts similarly to a rigid obstacle hindering local volume
expansion and, consequently, lithiation as discussed in Ref.~\cite{schoof2023simulation}.

We depict the evolution of the lithium concentration in
\cref{fig:c_stiff_sei}(e) to estimate the robustness of the mechanical impact
during cycling. During the first lithiation, the decrease in concentration at
point $\plr$ exists in the whole SOC range. During the subsequent delithiation,
this decrease in concentration reduces continuously, and an increase in
concentration, meaning an anomaly, appears for SOC values smaller than $20\%$.
During the second lithiation, a concentration anomaly appears for SOC values
larger than $30\%$. The second lithiation deviates significantly from the first
lithiation due to the viscoplastic behavior of the SEI shell. Nevertheless, the
concentration anomaly caused by the mechanical impact of the stiff SEI shell is
a robust effect appearing during every cycle.

The stiff SEI mechanics influences the chemo-mechanical potential inside
silicon. Due to the viscoplastic behavior, the stiff SEI shell generates a
stress hysteresis during cycling, causing a voltage hysteresis as depicted in
\cref{fig:voltage_soc_C20}. Thus, the hysteresis effect discussed in Refs.~\cite{kobbing2024voltage,kobbing2024slow} for a spherical silicon particle covered by a stiff
SEI shell also occurs for elliptical nanowires. This demonstrates the
importance of mechanical considerations for silicon cores and SEI shells in simulations dealing with silicon anodes as battery active material.



\section{Summary and Conclusion}
\label{sec:conclusion}

In this study, we have systematically
investigated the mechanical behavior and lithiation characteristics
of an elliptical silicon nanowire core covered by a viscoplastic SEI shell with
a 2D chemo-mechanical simulation. We have compared the influence of a soft and
stiff SEI shell on the system and discussed the effect of the elliptical
geometry. We base our model and numerical simulation on a higher order finite
element method with a variable-step, variable-order time integration
scheme extended straightforwardly from the 1D radial symmetric
case \cite{schoof2024comparison}.

Concerning the mechanics, the silicon and soft SEI system shows the largest
stress magnitudes in tangential direction at the end of the minor half-axis at
point $\pul$, where the curvature is minor. The normal component of the stress
shows the largest magnitude at the end of the major half-axis at point $\plr$,
however, with significantly smaller values compared to the tangential stresses.
For the stiff SEI case, the system reaches the largest stress magnitudes at the
end of the major half-axis at point $\plr$, where the curvature is major and
the mechanical impact of the SEI is dominant. Thus, the stress magnitudes are
significantly higher compared to the soft SEI case. Only the tangential stress
component inside the SEI is larger at the point with the smallest curvature
$\pul$, where the SEI is prone to cracking.
Symmetric silicon nanowires with the same capacity and corresponding SEI shell
are mechanically more stable than elliptical nanowires.

The mechanics of the elliptical geometry significantly influences the lithiation
behavior of the silicon nanowire. Generally, the lithium concentration is
increased at the outer boundary during lithiation and decreased during
delithiation, with the fastest concentration changes at the end of the major
half-axis at point $\plr$ due to the largest surface-to-volume ratio at this
point. For the soft SEI case, the concentration distribution reveals a
deviation from this trend at the end of the minor half-axis at point $\pul$.
This concentration anomaly also appears during slow cycling and without SEI.
Therefore, the mechanics of the elliptical silicon nanowire causes this effect.
For the stiff SEI case,
in contrast,
a concentration anomaly occurs at the end of the major half-axis at point
$\plr$. The SEI influences the lithiation behavior more dominantly at this
point due to the pronounced curvature. In total, the soft SEI has only a minor
effect on the silicon nanowire, while the stiff SEI significantly impacts the
lithiation behavior.

As shown in
Refs.~\cite{traskunov2021localized, traskunov2021new,
traskunov2022novel}, inhomogeneous lithiation on particle scale
is also responsible for considerable overpotential fluctuations
on electrode scale. Our results demonstrate that inclusion of
mechanical effects  not only predicts mechanical degradation
but also influences electrochemically induced degradation due
to the mechanically induced overpotential fluctuations.

To conclude, we have demonstrated the importance of the chemo-mechanical coupling, the geometry, and the SEI on the silicon anode behavior during cycling. Based on our work, further simulations could include plasticity of the silicon nanowire, fracture modes inside silicon and SEI, or SEI growth. From a numerical perspective, an adaptive spatial grid algorithm could optimize the simulation.



\section*{Acknowledgements}

\noindent
The authors thank L. von Kolzenberg
for fruitful discussions about modeling silicon particles
and G. F. Castelli for the software basis
as well as A. Dyck, J. Niermann and T. Böhlke for their contribution on
efficient (visco-) plastic modeling.
R.S. and L.K. acknowledge financial support by the German Research
Foundation~(DFG) through the Research Training Group 2218
SiMET~--~Simulation of Mechano-Electro-Thermal processes in Lithium-ion
Batteries, project number 281041241.
L.K. and B.H. acknowledge financial support by the European Union’s Horizon Europe
within the research initiative Battery 2030+ via the OPINCHARGE project under
the grant agreement number 101104032.
The authors acknowledge support by the state of Baden-Württemberg through bwHPC.

\section*{Conflict of Interest}

The authors declare that they have no known competing financial interests or
personal relationships that could have appeared to influence the work in this
paper.

\begin{shaded}
\noindent\textsf{\textbf{Keywords:} \keywords}
\end{shaded}

%
\section*{CRediT authorship contribution statement}

\noindent
\textbf{R. Schoof:}
Conceptualization, Methodology, Software, Validation, Formal analysis,
Investigation, Data Curation, Writing -- Original Draft, Visualization.
\textbf{L. Köbbing:}
Conceptualization, Methodology, Validation, Formal analysis,
Investigation, Writing -- Original Draft, Visualization.
\textbf{A. Latz:}
Resources, Writing -- Review \& Editing, Project administration,
Funding acquisition.
\textbf{B. Horst\-mann:}
Resources, Writing -- Review \& Editing, Supervision.
\textbf{W. Dörfler:}
Resources, Writing -- Review \& Editing, Supervision, Project administration,
Funding acquisition.

\section*{ORCID}
\noindent
R. Schoof:
\href{https://orcid.org/0000-0001-6848-3844}{0000-0001-6848-3844},
L. Köbbing:
\href{https://orcid.org/0000-0002-1806-6732}{0000-0002-1806-6732},
A. Latz: \href{https://orcid.org/0000-0003-1449-8172}{0000-0003-1449-8172},
B. Horstmann:
\href{https://orcid.org/0000-0002-1500-0578}{0000-0002-1500-0578},
W. Dörfler: \href{https://orcid.org/0000-0003-1558-9236}{0000-0003-1558-9236}

\renewcommand*{\bibfont}{\small}
\setlength\bibitemsep{0pt}
\printbibliography

\onecolumn
\appendix

\appendix

\addcontentsline{toc}{section}{Appendices}
\section*{Appendices}

\renewcommand{\thesection}{A}
\section{Symmetric Nanowire with Soft SEI}
\label{app:symm_si_soft_sei}

For comparison, we investigate the stresses and the lithiation characteristics
of a symmetric silicon nanowire during cycling. We choose the radius as
$L_\ptcl=\SI{38.73e-9}{\meter}$
to obtain the same capacity as for the elliptical nanowire and the thickness of the SEI is chosen as an eighth of the core length as $L_\sei=\SI{4.84e-9}{\meter}$.

\begin{figure}[!b]
	\centering
	\includegraphics[width = 0.8\textwidth,
	page=1]{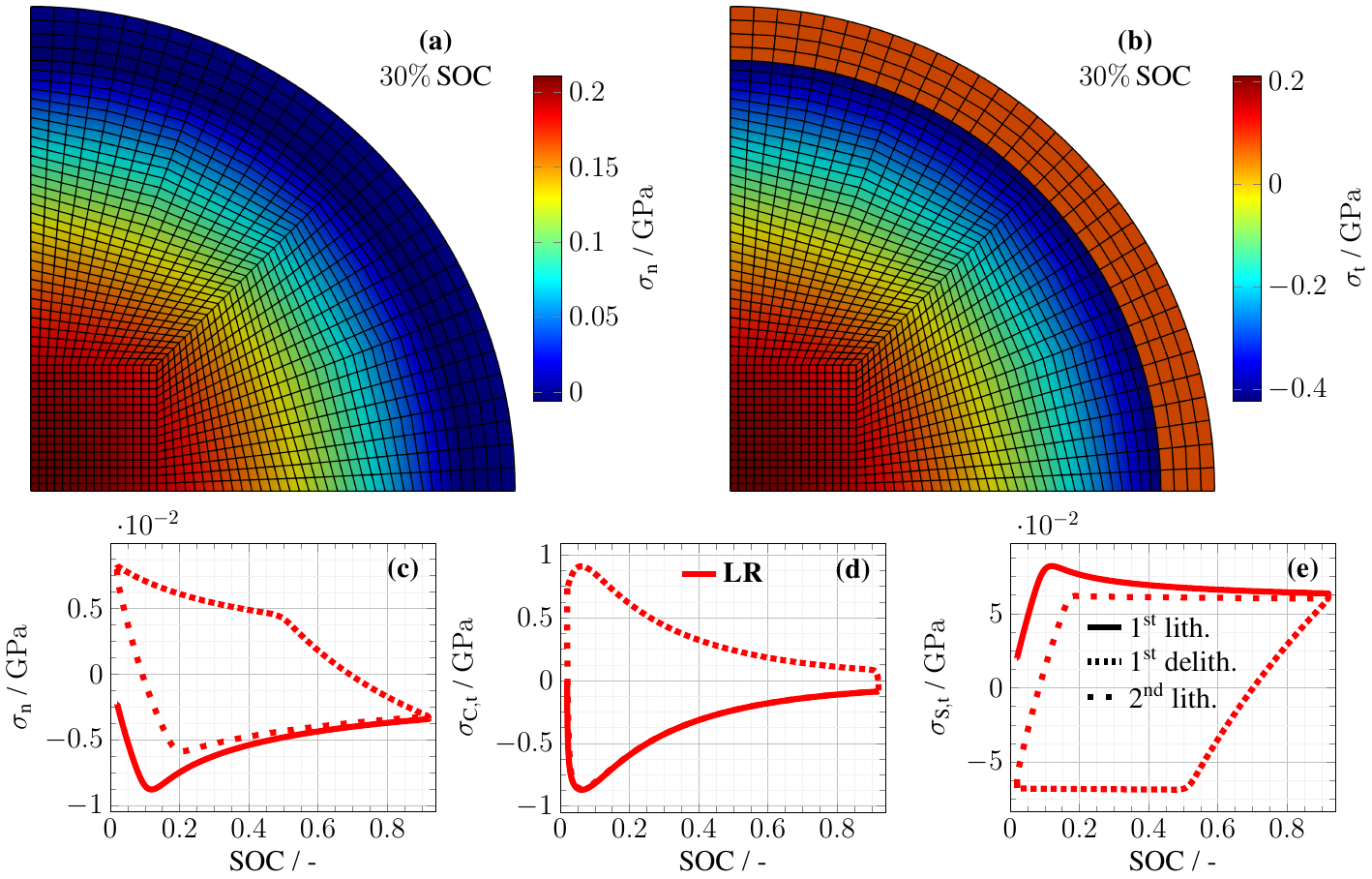}
	\caption{Cauchy stresses for the symmetric silicon nanowire with a soft SEI.
	Distribution of (a) normal and (b) tangential Cauchy stress inside the
	silicon core and the SEI shell during lithiation at $30\%$ SOC. Evolution of
	Cauchy stress at the interface during three half cycles for (c) normal, (d)
	tangential core, and (e) tangential shell stress.}
	\label{fig:sigma_symmetric}
\end{figure}

We depict the stress distribution during lithiation at $30\%$ SOC for the normal and tangential component in \cref{fig:sigma_symmetric}(a) and (b). Inside the silicon core, both components show compressive stresses close to the outer boundary and tensile stresses close to the center. Inside the SEI, the normal stress is compressive close to the nanowire and vanishes at the outer boundary. The tangential component is tensile inside the whole SEI shell.

\begin{figure}[!b]
	\centering
	\includegraphics[width = 0.8\textwidth,
	page=1]{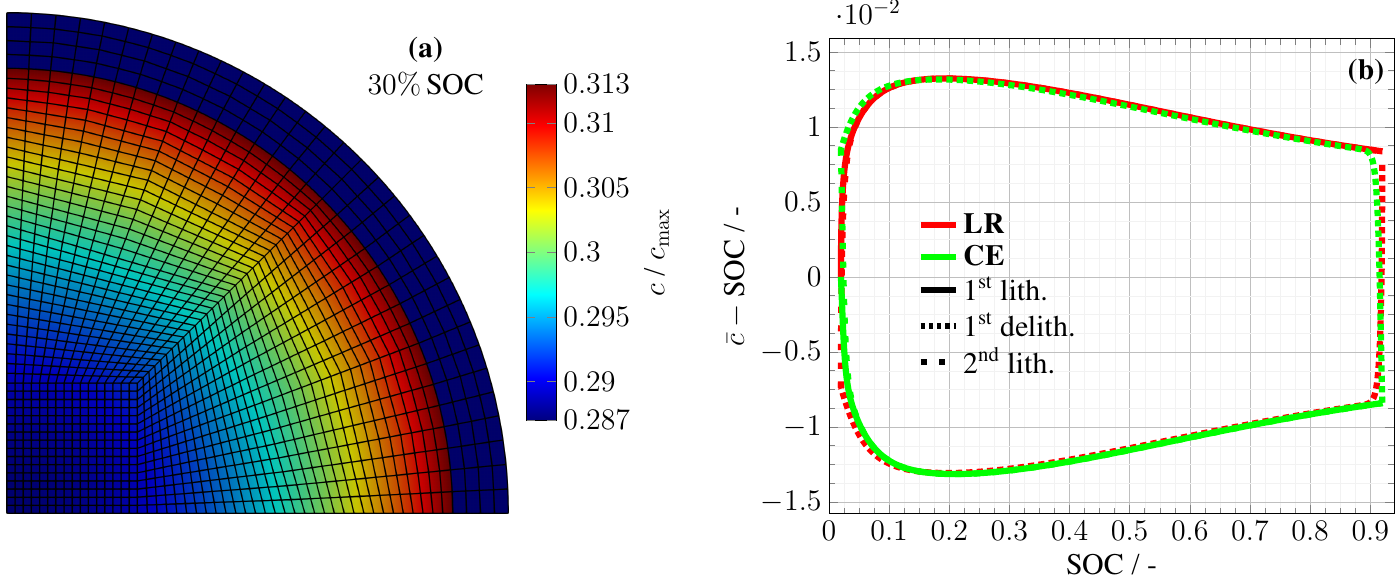}
	\caption{Lithium concentration for the symmetric silicon nanowire with a soft
	SEI. (a) Distribution of the lithium concentration inside silicon during
	lithiation at $30\%$ SOC. (b) Deviation of the lithium concentration from the
	mean at the interface $\plr$ and the center $\pc$ during three half cycles.}
	\label{fig:c_symmetric}
\end{figure}

We show the evolution of the stress components at the interface during three
half-cycles in \cref{fig:sigma_symmetric}(c), (d), and (e). During lithiation,
the normal component and the tangential component inside the silicon core show
compressive stress, while the tangential component inside the SEI shell shows
tensile stress. Stresses are opposite during delithiation. During the second
lithiation, the normal stress inside the silicon and the tangential stress
inside the SEI converge gradually to the stress during the first lithiation.
The tangential stress inside the silicon during the second lithiation coincides
with the first lithiation for the whole SOC range. The stress magnitude is
always smaller than the maximum magnitude in the elliptical case.

In \cref{fig:c_symmetric}(a), we depict the concentration distribution within
the symmetric silicon nanowire during lithiation at $30\%$ SOC. The
concentration increases from the outer boundary during lithiation and no
concentration anomaly occurs. The deviation of the lithium concentration at the
outer boundary and the center from the mean during cycling is shown in
\cref{fig:c_symmetric}(b). As expected, during lithiation, the lithium
concentration at the outer boundary exceeds the mean, while the concentration
at the center is smaller than the mean. The concentration profiles are vice
versa during delithiation, as expected.

\renewcommand{\thesection}{B}
\section{Variation of Plastic Strain Rate~$\dot{\varepsilon}_0$}
\label{app:different_dot_eps_0}

\begin{figure}[h]
  \centering
  \includegraphics[width = 0.8\textwidth,
  page=1]{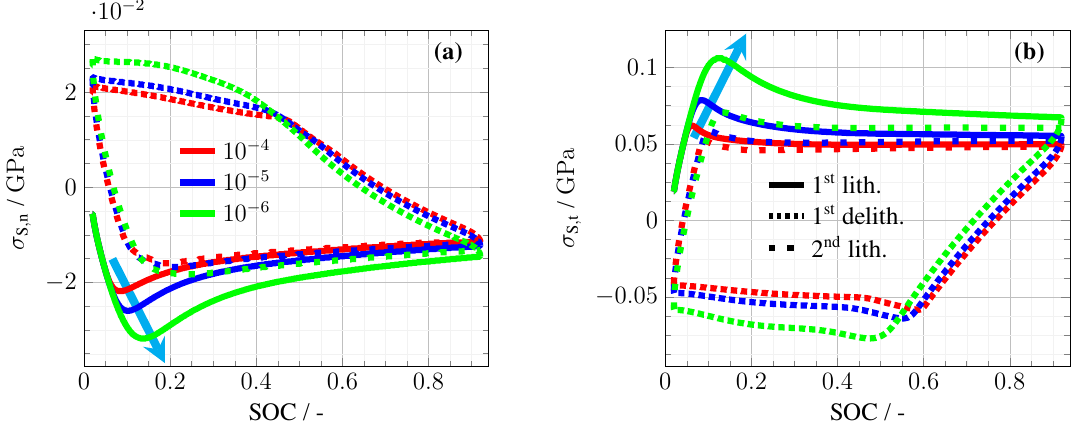}
  \caption{Variation of the plastic strain rate $\dot{\varepsilon}_0$ for the soft SEI. Evolution of the Cauchy stress inside the SEI during cycling for (a) the normal and (b) the tangential stress component.}
  \label{fig:different_dot_eps_0_all}
\end{figure}

To estimate the influence of the viscoplastic behavior, we vary the plastic
strain rate $\dot{\varepsilon}_0$ for the soft SEI. For higher values, plastic
flow starts quickly upon reaching the yield condition, while it starts only
slowly for smaller values. We depict the normal component of the stress inside
the SEI shell in \cref{fig:different_dot_eps_0_all}(a) and the tangential
component in \cref{fig:different_dot_eps_0_all}(b). The magnitude of both
stress components and the size of the stress overshoot increase with decreasing
plastic strain rate $\dot{\varepsilon}_0$. This is expected due to the retarded
plastic flow for low plastic strain rates $\dot{\varepsilon}_0$. Nevertheless,
the stress profiles reveal a similar shape for all tested values. For our
simulations, we take the medium parameter
$\dot{\varepsilon}_0=\SI{e-5}{\per\second}$.

\renewcommand{\thesection}{C}
\section{Stress-Driven Lithium Flux $\vect{N}_\el$}
\label{app:stress_driven_lithium_flux}

\begin{figure}[!b]
  \centering
  \includegraphics[width = 0.95\textwidth,
  page=1]{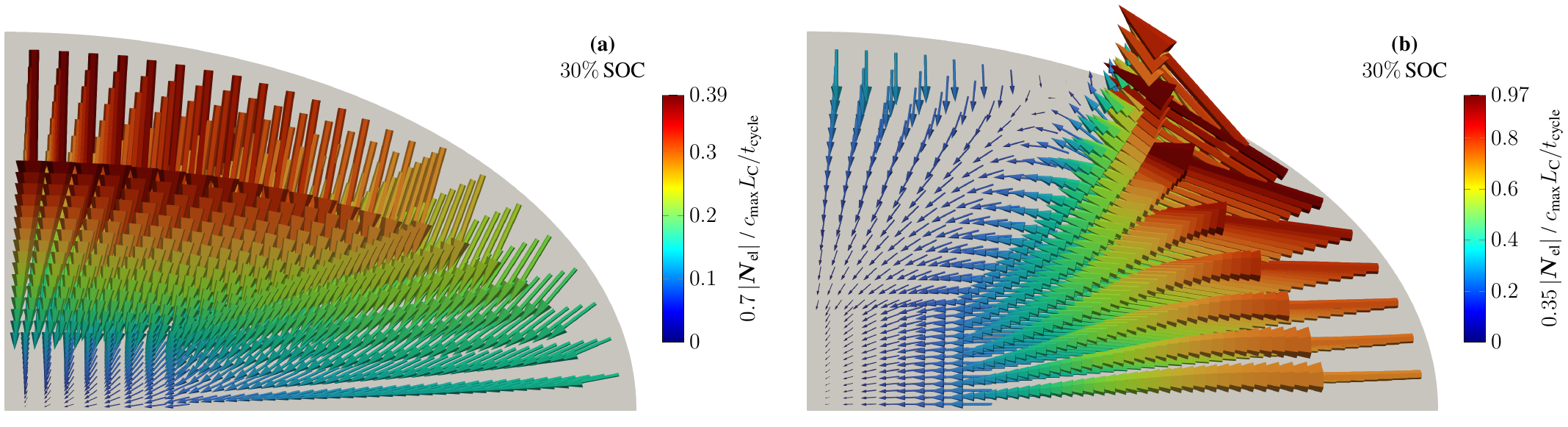}
  \caption{
      Stress-driven convective lithium flux $\vect{N}_\el$ for (a) the soft SEI
      case and (b) the stiff SEI case with different
      scaling for the fluxes.
    }
  \label{fig:stress_driven_lithium_flux}
\end{figure}

We depict the stress-driven convective lithium flux
$\vect{N}_\el$ during
lithiation at $30\%$ SOC in \cref{fig:stress_driven_lithium_flux} to complement
the illustrations of the concentration-driven diffusive lithium
flux~$\vect{N}_\ch$ in
\cref{fig:c_soft_sei}(d) and \cref{fig:c_stiff_sei}(d). For the soft SEI case
shown in \cref{fig:stress_driven_lithium_flux}(a), the stress-driven lithium
flux everywhere points towards the interior of the silicon core. The largest
magnitude of the stress-driven flux occurs at point $\pul$, where the largest
stress values and stress gradients exist. For the stiff SEI case shown in
\cref{fig:stress_driven_lithium_flux}(b), the stress-driven lithium flux mostly
points towards the interior of the silicon core. The largest magnitude of the
stress-driven flux and a significant deviation from center-directed flux occurs
at the
outer boundary in a larger region around point $\plr$, where the largest stress
values and stress gradients exist. Combining both flux components, the total
lithium flux always points towards the interior of the silicon core during
lithiation for the soft SEI as well as for the stiff SEI.

\newpage

\renewcommand{\thesection}{D}
\section{Silicon Nanowire without SEI and \si{C}/20}
\label{app:silicon_without_sei_C20}

\begin{figure}[!b]
	\centering
	\includegraphics[width = 0.95\textwidth,
	page=1]{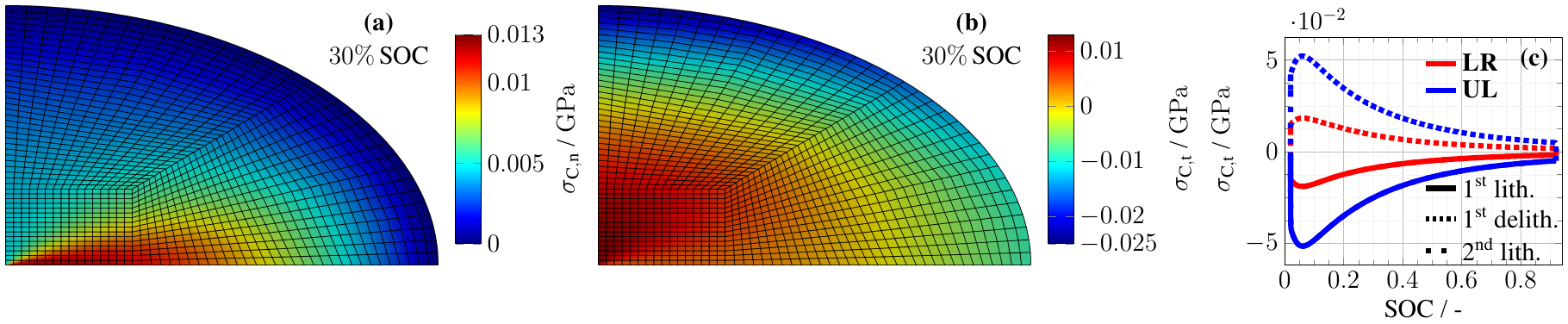}
	\caption{Cauchy stresses for the elliptical silicon nanowire without SEI
		during cycling with $\si{C}/20$. Distribution of (a) normal and (b)
		tangential Cauchy stress inside the silicon core during lithiation at $30\%$
		SOC. (c) Evolution of the tangential Cauchy stress at the points~$\plr$ and
		$\pul$ during three half cycles.}
	\label{fig:sigma_particle_only_C20}
\end{figure}

We briefly investigate the stress and lithiation characteristics of an
elliptical silicon nanowire without SEI during slow cycling with $\si{C}/20$ to
estimate the influence of the SEI and the $\si{C}$-rate.
The stress
distribution
in normal direction during lithiation at $30\%$ SOC is depicted in
\cref{fig:sigma_particle_only_C20}(a). The normal component vanishes at the
outer boundary due to the surface condition and shows tensile stress throughout
the nanowire. The largest stress magnitude is achieved close to the center
along the major half-axis. We display the tangential stress component in
\cref{fig:sigma_particle_only_C20}(b). The tangential component shows
compressive stress at the outer boundary, with the largest magnitude occurring
at the end of the minor half-axis at point $\pul$. The tangential stress is
tensile in a region around the center. In
\cref{fig:sigma_particle_only_C20}(c), we depict the evolution of the
tangential Cauchy stress during cycling at the points $\plr$ and $\pul$. The
curves reveal compressive stress during lithiation and tensile stress during
delithiation. The largest stress magnitudes occur at the end of the minor
half-axis at point $\pul$ for the whole SOC range. The stress distribution and
evolution during slow cycling is similar to the soft SEI case and cycling with
$1\,\si{C}$.
Only the magnitude of the stresses is smaller in general due to the
reduced $\si{C}$-rate.

\begin{figure*}[!b]
  \centering
  \includegraphics[width = 0.95\textwidth,
  page=1]{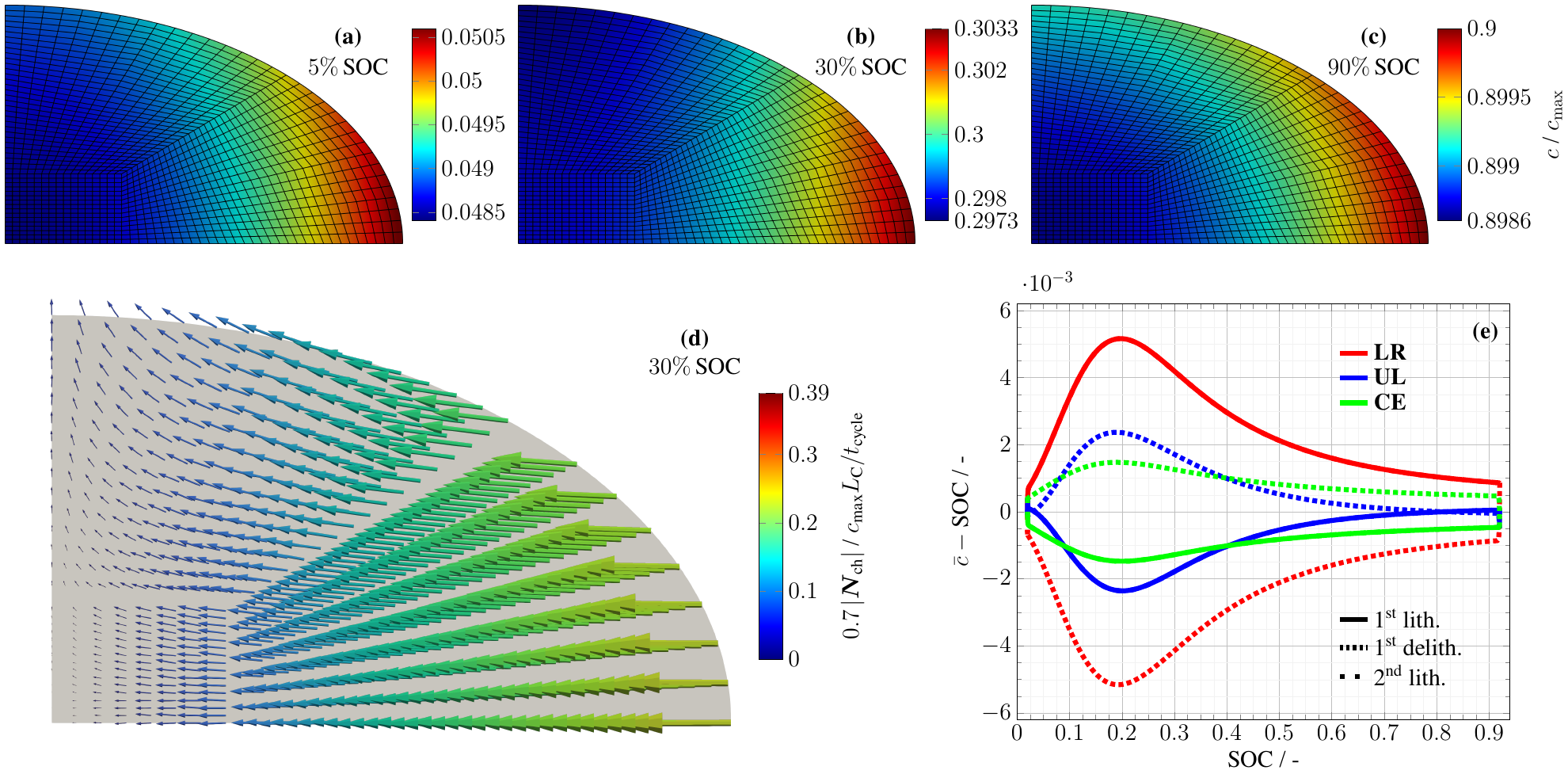}
  \caption{Concentration for silicon nanowire only with \si{C}/20.
  Lithium concentration for the elliptical silicon nanowire without SEI during
  cycling with $\si{C}/20$. Distribution of the lithium concentration inside
  silicon during lithiation at (a) $5\%$, (b) $30\%$, and (c) $90\%$ SOC.
  (d)~Concentration-driven diffusive lithium flux
  $\vect{N}_\ch$ scaled with $0.7$.
  (e)
  Deviation of the lithium concentration from the mean at the points~$\plr$, $\pul$, and
  $\pc$ during three half cycles.}
  \label{fig:c_particle_only_C20}
\end{figure*}

In \cref{fig:c_particle_only_C20}, we display
the lithium concentration
distribution during lithiation at (a) $5\%$, (b) $30\%$, and (c) $90\%$ SOC.
The concentration shows the largest values at the end of the major half-axis at
point $\plr$ due to the highest surface-to-volume ratio at this point. At
$30\%$ SOC, the distribution shows a concentration depletion at the end of the
minor half-axis at point $\pul$. The negatively scaled concentration gradient
during lithiation at $30\%$ SOC in \cref{fig:c_particle_only_C20}(d)
indicates the chemical diffusion component of
the lithium flux and
illustrates the
anomaly at point $\pul$. In \cref{fig:c_particle_only_C20}(e), we depict the
deviation of the lithium concentration at the points $\plr$, $\pul$, and $\pc$
from the mean during cycling. The evolution reveals the pronounced
concentration increase/decrease at point $\plr$ during lithiation/de\-lithiation.
Furthermore, the evolution reveals the concentration anomaly at point $\pul$
during cycling between $10\%$ and $40\%$ SOC. In total, the lithiation
characteristics during slow cycling without SEI are similar to standard
cycling
with SEI. Only the magnitude of the concentration deviations is significantly
reduced due to the smaller $\si{C}$-rate.

\renewcommand{\thesection}{E}
\section{Voltage Hysteresis}
\label{app:voltage_hysteresis}

\begin{figure*}[h]
	\centering
	\includegraphics[width = 0.75\textwidth,
	page=1]{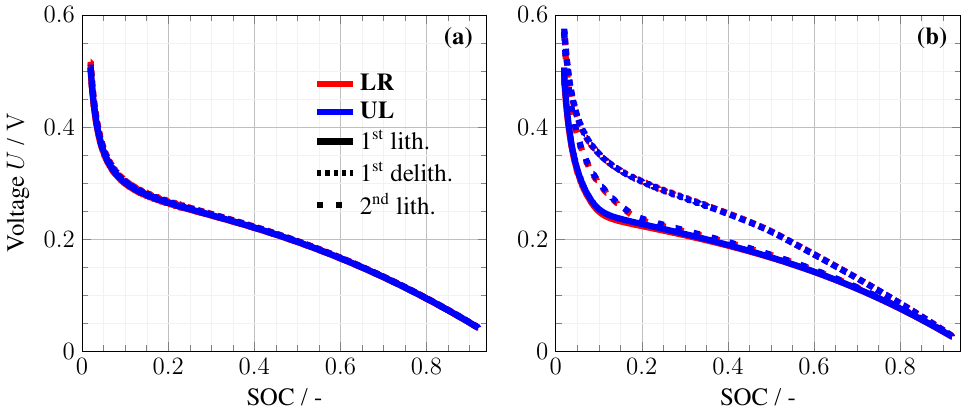}
	\caption{Voltage for the elliptical silicon nanowire covered by SEI during slow cycling with \si{C}/20 for (a) soft and (b) stiff SEI.}
	\label{fig:voltage_soc_C20}
\end{figure*}

To estimate the mechanical impact of the SEI shell on the lithiation behavior
of the silicon nanowire, we compare the voltage during slow cycling with
$\si{C}/20$ with the soft and the stiff SEI shell. Due to numerical reasons, we
adjust the plastic strain rate to
$\dot{\varepsilon}_0=\SI{e-6}{\per\second}$.
For both cases depicted in \cref{fig:voltage_soc_C20}, the voltages at point
$\plr$ and $\pul$ are equivalent, revealing chemo-mechanical equilibrium during
slow cycling. For the soft SEI case displayed
in \cref{fig:voltage_soc_C20}(a),
also the voltages during lithiation and delithiation coincide. In contrast,
\cref{fig:voltage_soc_C20}(b) reveals that a voltage hysteresis arises for the
stiff SEI case. This is in agreement with our explanation of the voltage
hysteresis for spherical silicon nanoparticles covered by a stiff SEI shell
presented in Refs.~\cite{kobbing2024voltage,kobbing2024slow}.

\renewcommand{\thesection}{F}
\section{Table with Parameters}
\label{app:sigma_t_sei}

The simulation parameters and constants are summarized in \cref{tab:parameters}. Additionally, we follow Ref.~\cite{kolzenberg2022chemo-mechanical} and use
$U_{\max} = \SI{0.5}{V}$
and $U_{\min} = \SI{0.05}{V}$
as maximal and minimal voltage for the lithiation and delithiation.
Therefore, we choose $\conb_0=0.02$ as constant initial concentration and
$\SI{0.9}{h}$ as duration of one half cycle.
The applied OCV curve
\begin{align}
  \label{eq:ocv}
  U_\text{OCV}(\conb)
  &
  =
  \frac{\minus0.2453 \, \conb^3-0.00527 \, \conb^2+0.2477 \, \conb+0.006457}
  {\conb+0.002493}
\end{align}
is delivered
by Ref.~\cite{chan2007high-performance}.

\begin{table}[!h]
  \centering
  \caption{Model parameters for the numerical experiments
    \cite{kolzenberg2022chemo-mechanical, schoof2024modeling,
      di-leo2015diffusion-deformation}.}
  \label{tab:parameters}
  \begin{tabular}[t]{@{}llccc@{}}
    \toprule
    \textbf{Description} & \textbf{Symbol} &
    \multicolumn{1}{c}{\textbf{Value}} & \textbf{Unit} &
    \multicolumn{1}{c}{\textbf{Dimensionless}} \\

    \midrule

    Universal gas constant & $R_\text{gas}$ & $8.314$ &
    \si{\joule\per\mol\per\kelvin}
    & $1$ \\[\defaultaddspace]

    Faraday constant & $F$ & $96485$ & \si{\joule\per\volt\per\mol} &
    $1$
    \\[\defaultaddspace]

    Operation temperature & $T$ & $298.15$ & \si{\kelvin} & $1$
    \\[\defaultaddspace]

    \midrule

    \multicolumn{5}{c}{Silicon} \\

    \midrule

    Core length scale & $L_\ptcl$ & $\num{50e-9}$ &
    \si{\meter} & 1
    \\[\defaultaddspace]

    Cycle time & $t_\text{cycle}$ & $\num{3600}$ &
    \si{\second} & 1
    \\[\defaultaddspace]

    Diffusion coefficient & $D$ & $\num{1e-17}$ &
    \si{\square\meter\per\second} &
    $14.4$ \\[\defaultaddspace]

    OCV curve & $U_\text{OCV}$ & $\text{Equation~}\eqref{eq:ocv}$ &
    \si{\volt}
    & $F/R_\text{gas} T \cdot \eqref{eq:ocv}$
    \\[\defaultaddspace]

    Young's modulus & $E_\ptcl$ & $\num{90.13e9}$ & \si{\pascal} & $116.74$
    \\[\defaultaddspace]

    Partial molar volume & $v_\text{pmv}$ & $\num{10.96e-6}$ &
    \si{\cubic\meter\per\mol} & $3.41$
    \\[\defaultaddspace]

    Maximal concentration & $c_\text{max}$ & $\num{311.47e3}$ &
    \si{\mol\per\cubic\meter} & $1$
    \\[\defaultaddspace]

    Initial concentration & $c_0$ & $\num{6.23e3}$ &
    \si{\mol\per\cubic\meter} & $\num{2e-2}$
    \\[\defaultaddspace]

    Poisson's ratio & $\nu_\ptcl$ & $0.22$ & \si{-} & $0.22$
    \\[\defaultaddspace]

    \midrule

    \multicolumn{5}{c}{SEI} \\

    \midrule

    Shell length scale & $L_\sei$ & $\num{6.25e-9}$ &
    \si{\meter} & 0.125
    \\[\defaultaddspace]

    Young's modulus & $E_\sei$ & $\num{900e6}$ & \si{\pascal} & $1.17$
    \\[\defaultaddspace]

    Poisson's ratio & $\nu_\sei$ & $0.25$ & \si{-} & $0.25$
    \\[\defaultaddspace]

    Yield stress & $\sigma_\text{Y}$ & $\num{49.5e6}$ &
    \si{\pascal} & $0.052$
    \\[\defaultaddspace]

    Strain measurement & $\beta$ & $\num{2.94}$ &
    \si{-} & $\num{2.94}$
    \\[\defaultaddspace]

    Stress constant & $\sigma_{\text{Y}^{*}}$ & $\num{49.5e6}$ &
    \si{\pascal} & $0.052$
    \\[\defaultaddspace]

    Tensile plastic strain rate & $\dot{\varepsilon}_0$ & $\num{1.0e-5}$ &
    \si{\per \second} & $0.036$
    \\

    \bottomrule
  \end{tabular}
\end{table}





\twocolumn
%
%




\end{document}